\documentclass[reprint,aps,prl,superscriptaddress]{revtex4-1}
\usepackage{graphicx}
\usepackage{hyperref}
\usepackage{amsmath}
\usepackage{amssymb}
\usepackage{dcolumn}
\usepackage{bm}
\usepackage{setspace}
\usepackage{color}
\usepackage[capitalize,nameinlink]{cleveref}

\begin{document}

\title{\LARGE Double-mode relaxation of highly deformed vesicles}

\author{Dinesh Kumar}
\affiliation
{
	Department of Chemical and Biomolecular Engineering \\ University of Illinois at Urbana-Champaign, Urbana, IL, 61801
}
\affiliation
{
	Beckman Institute for Advanced Science and Technology \\ University of Illinois at Urbana-Champaign, Urbana, IL, 61801
}

\author{Channing M. Richter}
\affiliation
{
	Department of Chemical and Biomolecular Engineering \\ University of Illinois at Urbana-Champaign, Urbana, IL, 61801
}

\author{Charles M. Schroeder}
\email[To whom correspondence must be addressed: ]{cms@illinois.edu}
\affiliation
{
	Department of Chemical and Biomolecular Engineering \\ University of Illinois at Urbana-Champaign, Urbana, IL, 61801
}
\affiliation
{
	Beckman Institute for Advanced Science and Technology \\ University of Illinois at Urbana-Champaign, Urbana, IL, 61801
}
\affiliation
{
	Department of Materials Science and Engineering \\ University of Illinois at Urbana-Champaign, Urbana, IL, 61801
}

\date{\today}

\begin{abstract}
Lipid vesicles are known to undergo complex conformational transitions, but it remains challenging to systematically characterize non-equilibrium membrane shape dynamics. Here, we report the direct observation of lipid vesicle relaxation from highly deformed shapes using a Stokes trap. Vesicle shape relaxation is described by two distinct characteristic time scales governed by the bending modulus and membrane tension. Interestingly, experimental results are consistent with a viscoelastic model of a deformed membrane. Overall, these results show that vesicle relaxation is governed by an interplay between membrane elastic moduli, surface tension, and vesicle deflation.
\end{abstract}

\maketitle

Membrane-bound vesicles are ubiquitous in biological systems \cite{mashburn2005membrane,sowinski2008membrane} and drug delivery applications \cite{langer1990new}. Phospholipid vesicles are often used as model systems to study the mechanical properties of living cells \cite{lipowsky1991conformation,noguchi2005shape,boal2002mechanics,fenz2012giant,chen2014large,keber2014topology,dimova2019giant}. In addition, synthetic vesicles serve as triggered-release agents \cite{amstad2011triggered} or encapsulants in detergents and fabric softeners \cite{jesorka2008liposomes}. In many cases, structure-property relations underlie the functional behavior of these materials. Despite recent progress, however, the non-equilibrium shape dynamics of vesicles is not yet fully understood \cite{misbah2006vacillating,danker2007rheology,kaoui2009red,danker2009vesicles,coupier2012shape,kantsler2006transition}. 

Vesicles undergo a wide array of stretching dynamics in flow depending on the flow type and equilibrium vesicle shape \cite{kantsler2005orientation,kantsler2008critical,kantsler2007vesicle,dahl2016experimental,kumar2020conformational,lin2019shape}. In shear flow, vesicles exhibit tumbling, tank-treading, and membrane trembling behavior that depends on the flow strength and viscosity ratio \cite{kantsler2005orientation,kantsler2006transition,deschamps2009phase}. In extensional flow, vesicles with non-spherical equilibrium shapes exhibit a wide array of conformational transitions, including a tubular-to-symmetric dumbbell transition for highly deflated vesicles \cite{kantsler2008critical,kumar2020conformational,narsimhan2014mechanism,narsimhan2015pearling}, a spheroidal-to-asymmetric dumbbell transition for moderately deflated vesicles \cite{narsimhan2014mechanism,dahl2016experimental,kumar2020conformational}, and a nearly spherical-to-ellipsoidal transition for weakly deflated vesicles \cite{lebedev2007dynamics,zhao2013shape,zhou2011stretching,wu2015viscoelastic,kumar2020conformational}. Such deformable membrane behavior is naturally exploited in biological systems; for example, red blood cells readily adopt biconcave disk shapes \cite{pivkin2008accurate}, enabling large reversible deformation while traversing thin capillaries during circulation. 

Vesicle relaxation following deformation is critically important for shape dynamics and reversible elastic behavior \cite{boal2002mechanics}. Prior work has focused on the near-equilibrium relaxation of quasi-spherical vesicles following small deformations, induced by relatively weak forces using optical tweezers \cite{zhou2011stretching} or electrodeformation \cite{yu2015ellipsoidal}. Kantsler \textit{et al.} \cite{kantsler2008critical} observed the relaxation of a weakly deformed tubular-shaped vesicle, albeit only for a small ensemble size. Broadly, fundamental studies of shape relaxation for freely suspended vesicles following large deformations are challenging due to the need for precise flow control and manipulation without using micropipettes or direct physical contact of membranes.

\begin{figure*}
\begin{center}
\includegraphics[width=1.0\textwidth]{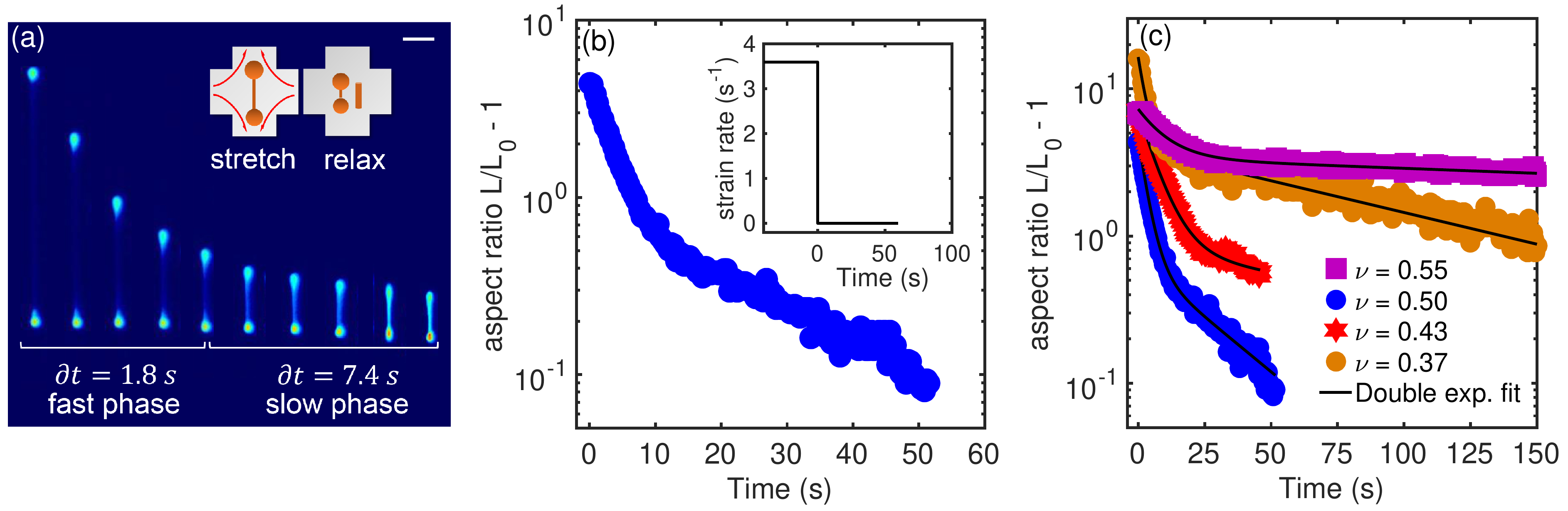}
\caption{\label{fig:relax_snapshot}Relaxation of highly deformed lipid vesicles following deformation in extensional flow using a Stokes trap. (a) Time-series of images showing a vesicle ($\nu=0.5$) relaxing back to an equilibrium tubular shape after being deformed into a symmetric dumbbell in extensional flow. Time between snapshots is $\delta t=1.8 s$ for fast phase and 7.4 s for the slow phase. Scale bar is 20 $\mu m$. (b) Time trajectory of vesicle shape relaxation following deformation in flow, showing semi-log plot of the aspect ratio $L/L_0-1$ as a function of time for the vesicle in \cref{fig:relax_snapshot}a. \textit{Inset}: Flow deformation protocol prior to vesicle relaxation. At time $t$=0, the flow is stopped and membrane relaxes back to equilibrium. (c) Semi-log plot of the aspect ratio $L/L_0-1$ as a function of time for different vesicles having reduced volume $\nu=0.55,0.5,0.43,0.37$ with corresponding double-exponential fits.}
\end{center}
\end{figure*}

Here, we report the direct observation of vesicle relaxation following large deformations in extensional flow. Vesicles with non-spherical shapes at equilibrium are deformed in precisely controlled flows using a Stokes trap \cite{Shenoy12042016,kumarorientation2019,shenoy2019flow}, followed by relaxation under quiescent conditions. Remarkably, our results show that highly deformed, freely suspended vesicles relax by a double-mode exponential pathway governed by two distinct and well-separated time scales corresponding to characteristic bending and surface tension time scales.

Giant unilamellar vesicles (GUVs) are prepared from a mixture of 1,2-dioleoyl-sn-glycero-3-phosphocholine (DOPC) and 0.12 mol \% of the fluorescent lipid 1,2-dioleoyl-sn-glycero-3-phosphoethanolamine-N-(lissamine rhodamine B sulfonyl (DOPE-Rh) in 100 mM sucrose buffer using an electroformation method (Supplemental Material) \cite{angelova1992preparation,kumar2020conformational}. Following electroformation, GUVs are slightly deflated by introducing a higher osmolarity sucrose solution to the outer fluid. Deflated vesicles are described by a reduced volume $\nu=(3V\sqrt{4\pi}) / A^{3/2}$, where $V$ and $A$ are the equilibrium vesicle volume and surface area, respectively, determined by revolution of the membrane contour, as previously described \cite{dahl2016experimental,kumar2020conformational}. In this way, $\nu$ is a measure of vesicle asphericity, such that $\nu$=1 corresponds to a perfectly spherical shape. 

Prior to vesicle stretching experiments, we determined the average bending modulus of quasi-spherical vesicles to be $\kappa = 22.3$ $k_BT$ using fluctuation spectroscopy, as previously described \cite{pecreaux2004refined,kumar2020conformational}. In all subsequent experiments, vesicles are deformed in the bending-dominated regime, such that no area stretching of the membrane occurs in the initial stretching step prior to vesicle relaxation (Supplemental Material). The cross-over tension from the bending to the area stretching regime is given by $\sigma = K_ak_BT / 8\pi\kappa \approx 0.1$ mN/m \cite{fournier2001effective,dimova2002hyperviscous}, where $K_a$ is the area-stretching modulus, $k_B$ is the Boltzmann constant, $T$ is absolute temperature, and $\kappa$ is the bending modulus. In our experiments, the maximum membrane tension is typically one order of magnitude smaller than the cross-over tension (Supplemental Material, Table S1). 

We began by studying the conformational relaxation of nearly spherical vesicles ($\nu > $ 0.9) using the Stokes trap (Supplemental Material, Fig. S1). Quasi-spherical vesicles ($\nu$ = 0.95) retain an ellipsoidal shape in extensional flow without transitioning into a dumbbell shape (Fig. S2) \cite{yu2015ellipsoidal,kumar2020conformational,zhou2011stretching,wu2015viscoelastic}. Following cessation of flow, nearly spherical vesicles undergo a rapid initial retraction (first few images in Fig. S2) that occurs at a rate slightly slower than the sampling rate of the imaging system (Supplemental Material). The rapid initial retraction is followed by a slow relaxation response that is well described by a single exponential decay, as shown in Fig. S3. Overall, these observations for quasi-spherical vesicles are consistent with prior electrodeformation experiments \cite{yu2015ellipsoidal} and theoretical predictions \cite{Seifert1999,liu2017deformation}. Based on these results, we hypothesized that floppy vesicles may exhibit a double-mode relaxation process that would show longer timescale dynamics compared to nearly spherical vesicles. 

We next characterized the transient relaxation dynamics of deflated vesicles with reduced volumes $\nu < $ 0.75 (\cref{fig:relax_snapshot}). Here, vesicles are deformed in extensional flow using strain rates $\dot{\epsilon}$ larger than the critical strain rate $\dot{\epsilon}_c$ required for shape deformation for floppy vesicles \cite{kumar2020conformational}. Using the Stokes trap, vesicles are deformed in flow for $\approx$40 s, thereby generating large accumulated fluid strains $\epsilon$ = $\dot{\epsilon}t$ = 40-400 (Supplemental Material, Fig. S1) and achieving vesicle aspect ratios $L/L_0 \approx$ 4-16, where $L$ is the vesicle stretch along the axis of extension and $L_0$ is the vesicle extension at equilibrium in the absence of flow. During the deformation step ($\dot{\epsilon} > \dot{\epsilon}_c$), a vesicle undergoes a non-equilibrium shape transition into a symmetric dumbbell for 0.25 $< \nu < 0.60$ (\cref{fig:relax_snapshot}a, Supplementary Movie 1) or an asymmetric dumbbell for 0.60 $< \nu < 0.75$ with a long, thin tether \cite{kumar2020conformational}. Following deformation, the flow field is abruptly stopped, and vesicles relax back to an equilibrium shape under quiescent conditions. Experiments are performed for vesicles located near the center-plane of the microdevice in the vertical direction, such that the strain rate is well characterized using particle tracking velocimetry (Supplemental Material, Fig. S4) \cite{kumar2020conformational}.

A characteristic transient relaxation trajectory for a highly deformed vesicle ($\nu$ = 0.5) is shown in \cref{fig:relax_snapshot}b, with the time series of images shown in \cref{fig:relax_snapshot}a. Prior to flow cessation, the vesicle deforms into a symmetric dumbbell with a long, thin tether connecting the two spherical ends (\cref{fig:relax_snapshot}a). Following flow cessation, the vesicle eventually relaxes back to its equilibrium shape. The transient relaxation trajectory reveals that vesicle shape relaxes via two stages: an initial fast retraction step, where the length of thin tether rapidly shortens, followed by a slow relaxation step in which the vesicle returns to an equilibrium shape.

\begin{figure}[t]
	\centering
	\includegraphics[width=0.4\textwidth]{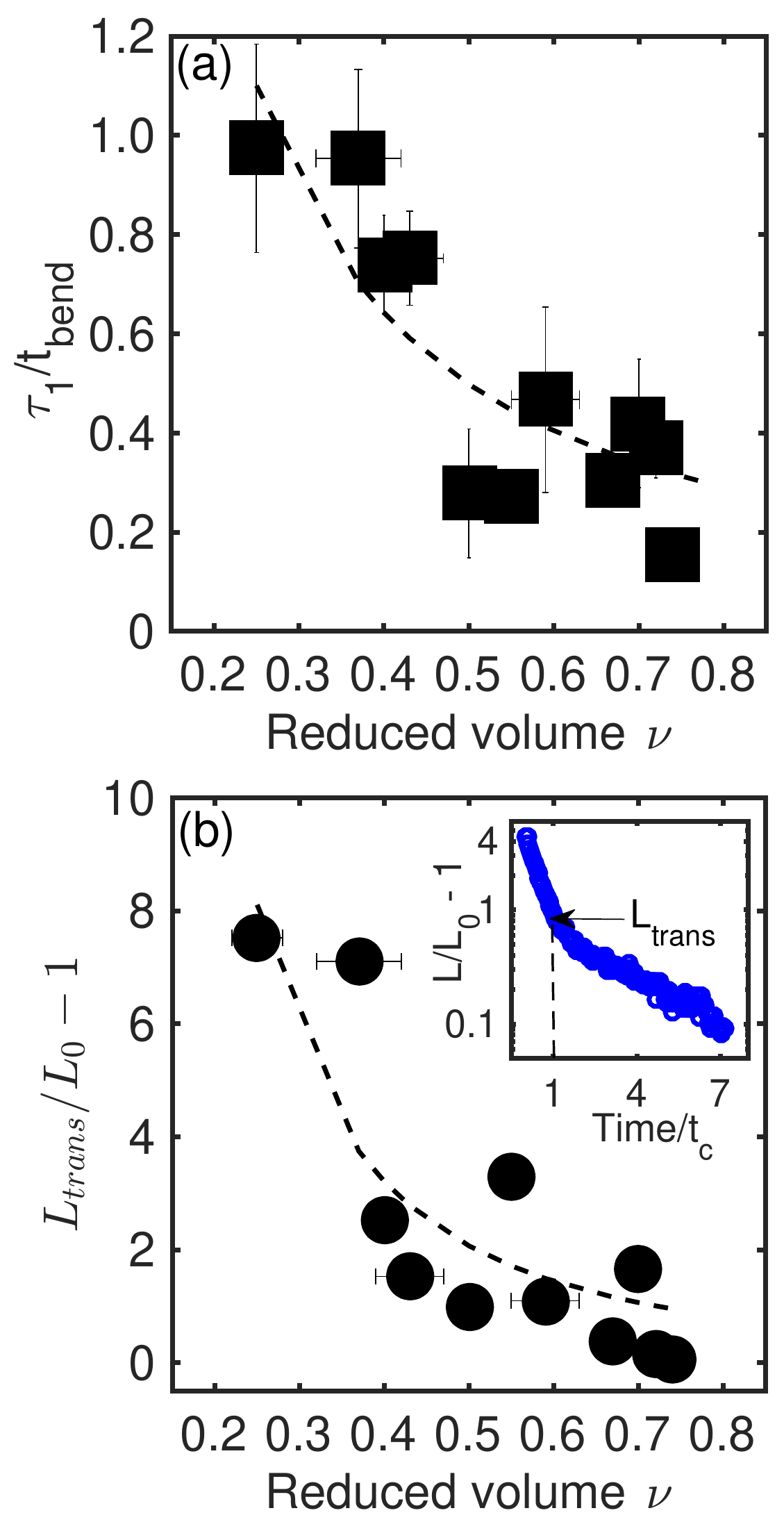}
	\caption{\label{fig:fast_tau1} Characterization of the fast double-mode relaxation time $\tau_1$. (a) Normalized time-scale $\tau_1/t_{bend}$ as a function of reduced volume $\nu$. (b) Transition aspect ratio  $L_{trans}/L_0-1$ at the cross-over time $t_c$ between the fast and slow relaxation phases as a function of reduced volume $\nu$. \textit{Inset}: A representative relaxation trajectory for $\nu=0.5$ showing $L_{trans}$ at cross-over time $t_c$.}
\end{figure}

A series of characteristic relaxation trajectories for vesicles with different reduced volumes $\nu$ is shown in \cref{fig:relax_snapshot}c (Supplemental Material, Fig. S5), where the vesicle aspect ratio is defined by $L(t)/L_0-1$, and $L(t)$ is the time-dependent vesicle extension along the elongational axis. In all cases, our results show that the relaxation trajectories can be described by a double-mode exponential decay:
\begin{align}
\frac{L(t)}{L_0}-1 = A \exp(-t/\tau_1) + B \exp(-t/\tau_2)
\label{eq:exp_fit}
\end{align}
where $\tau_1$ and $\tau_2$ are the fast and slow relaxation times, respectively, and $A$ and $B$ are numerical constants. Vesicle relaxation trajectories are well described by a double-mode exponential function across a wide range of reduced volumes, as shown in \cref{fig:relax_snapshot}c. Repeated relaxation experiments on the same vesicle show nearly identical relaxation trajectories (Supplemental Material, Fig. S6), which is consistent with the notion that the ratio of thermal energy to the bending modulus $k_BT/\kappa \approx 0.04$ is small, suggesting that vesicle shape relaxation governed by membrane bending fluctuations is nearly deterministic under repeated relaxation trials. 

We determined the double-mode relaxation times $\tau_1$ and $\tau_2$ for vesicles over a wide range of reduced volumes (0.25 $< \nu <$ 0.75), as shown in \cref{fig:relax_snapshot}c (Supplemental Material, Fig. S5, Table S1). Interestingly, the numerical values of the fast relaxation time $\tau_1$ are on the order of the characteristic bending relaxation time $t_{bend} = \mu R^3/\kappa$, where $\mu$ is the viscosity of suspending medium and $R$ is the equivalent radius of a vesicle determined from $R=\sqrt{(A/4\pi)}$ (\cref{fig:fast_tau1}a). Here, the time scale $t_{bend}$ corresponds to the leading order time constant for the longest bending mode, but other fluctuation modes exist in the full spectrum \cite{PhysRevA.36.4371}. These results suggest that $\tau_1$ corresponds to relaxation of long-wavelength bending modes following a non-linear membrane deformation. Interestingly, repeated relaxation trials on the same vesicle show that $\tau_1$ does not depend on the strain rate used in the deformation step (Supplemental Material, Table S2, Fig. S7), which is consistent with the notion that vesicles are deformed in the bending-dominated regime with no significant changes in the membrane structure. 

Unexpectedly, our data show that the normalized fast double-mode relaxation time $\tau_1/t_{bend}$ is a weak function of the vesicle reduced volume $\nu$, as shown in \cref{fig:fast_tau1}a. In particular, the normalized fast retraction time $\tau_1/t_{bend}$ decreases as the reduced volume increases, which is consistent with the fact that vesicles with smaller reduced volumes have larger surface area to volume ratios and hence larger degrees of membrane floppiness. In \cref{fig:fast_tau1}, error bars for reduced volume arise from measurement uncertainty in the vesicle equivalent radius $R$, propagated from vesicle surface area $A$ and volume $V$, as previously described \cite{kumar2020conformational}.

Following the initial fast relaxation step, the vesicle membrane transitions to a slow relaxation process described by a second time scale $\tau_2$. Interestingly, the numerical values of $\tau_2$ are on the order of the characteristic surface tension time scale $t_{surf} = \mu L_{trans}/\sigma_0$ (Supplemental Material, Fig. S8), where $\sigma_0$ is the ensemble-averaged equilibrium tension and $L_{trans}$ is the vesicle stretch at the cross-over time between the fast and slow regimes, defined as $t_c = C \tau_1\tau_2 /(\tau_2-\tau_1)$ from \cref{eq:exp_fit}, where $C = \ln B/A$ \cite{kumar2020conformational}. In particular, our results show that $\tau_2$ is within an order of magnitude of $t_{surf}$, which is consistent with the relatively broad distribution of membrane tensions known to result from generating vesicles using electroformation \cite{yu2015ellipsoidal,kumar2020conformational,dahl2016experimental}. Here, the membrane tension for any single vesicle may vary by an order of magnitude around the mean value $\sigma_0$ for the ensemble. Overall, the slow mode described by $t_{surf}$ is analogous to the relaxation of Newtonian fluid drops following deformation in flow, where a constant surface tension drives the drop back to its equilibrium spherical shape \cite{ha2001experimental}. 

\begin{figure}[t]
	\includegraphics[width=.40\textwidth]{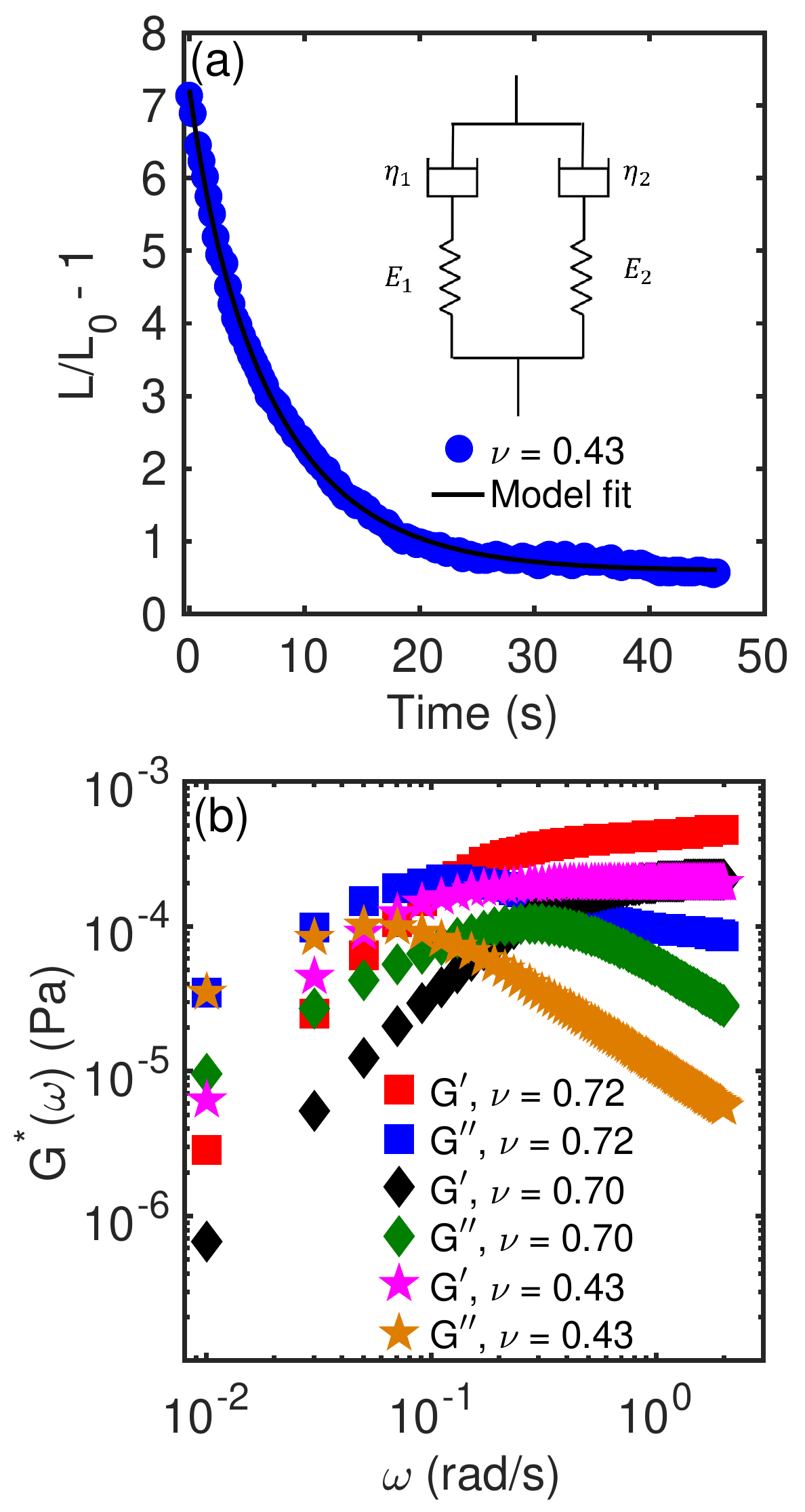}
	\caption{\label{G_complex} Viscoelastic model for vesicle membranes. (a) Fit of viscoelastic model to representative relaxation trajectory for vesicle with reduced volume $\nu=0.43$. (b) Viscoelastic model predictions of storage $G^*(\omega)$ and loss $G^{\prime\prime}(\omega)$ moduli for three vesicles with different reduced volumes.}
\end{figure}

Interestingly, our results further show that the vesicle aspect ratio at the transition between the fast and slow relaxation phases ($L_{trans}/L_0-1$) is a weak function of reduced volume (\cref{fig:fast_tau1}b). These results suggest that the bending modulus $\kappa$ may have a functional dependence on the reduced volume $\nu$. Broadly, these findings might have origins in more subtle aspects of membrane mechanics. Deflated vesicles are known to exhibit an imbalance in lipid density in the two leaflets of the membrane that induces a spontaneous curvature generation \cite{PhysRevE.49.5389}. Such modification of the spontaneous curvature and high lateral diffusion of individual lipid molecules in the bilayer for vesicles with smaller reduced volumes may reduce the energy requirements for thermal fluctuations, thereby resulting in a dependence of bending modulus on the degree of osmotic deflation \cite{fa2007decrease}. Finally, relaxation of highly deformed vesicles with a dumbbell shape involves fluid flow through the thin tether connecting the bulbs. Resistance to flow through the thin tether introduces an additional time scale $\mu L_{max}^2R_v / \sigma_0 r_t^2$, where $R_v$ is the radius of spherical bulb consuming the thin tether, and $r_t$ is tether radius at the beginning of relaxation. For our experiments, this time scale is several orders of magnitude larger than $\tau_1$ and $\tau_2$, suggesting that bending fluctuations and surface tension drive the conformation relaxation of deformed vesicles. 

The double-mode relaxation behavior of a vesicle membrane can be described by a viscoelastic model consisting of two Maxwell elements in a parallel arrangement with moduli ($E_1, E_2$) and viscosities ($\eta_1, \eta_2$), (\cref{G_complex}, Supplemental Material). Experimental data on vesicle relaxation is well described by the two-mode viscoelastic model, as shown in \cref{G_complex}a, thereby enabling determination of the model parameters $E_1,E_2, \eta_1$ and $\eta_2$. Using the stress relaxation model and parameters determined by fitting to experimental data, the frequency-dependent complex shear modulus $G^*(\omega)$ can be determined. In particular, the storage modulus $G^\prime(\omega)$ and the loss modulus $G^{\prime\prime}(\omega)$ can be obtained for single vesicles (Supplemental Material) \cite{mason1995optical}: 
\begin{align}
{G}'(\omega)=\frac{E_{1}\left ( \omega \frac{\eta_{1}}{E_{1}} \right )^2}{1+\left ( \omega \frac{\eta_{1}}{E_{1}} \right )^2}+\frac{E_{2}\left ( \omega \frac{\eta_{2}}{E_{2}} \right )^2}{1+\left ( \omega \frac{\eta_{2}}{E_{2}} \right )^2}\\
{G}''(\omega)=\frac{E_{1} \omega \frac{\eta_{1}}{E_{1}} }{1+\left ( \omega \frac{\eta_{1}}{E_{1}} \right )^2}+\frac{E_{2} \omega \frac{\eta_{2}}{E_{2}} }{1+\left ( \omega \frac{\eta_{2}}{E_{2}} \right )^2}
\end{align}
where $\omega$ is the deformation frequency. 

Plots of $G^\prime(\omega)$ and $G^{\prime\prime}(\omega)$ for vesicles with different reduced volumes are shown in \cref{G_complex}b. Results from model predictions show that the elastic modulus $G^\prime(\omega)$ increases with frequency and becomes larger than the viscous modulus $G^{\prime\prime}(\omega)$ at a cross-over frequency corresponding to the fast-time scale $\tau_1$, which is a signature of a transition from fluid to solid-like behavior. Moreover, the model predicts an approximate plateau modulus $G_0 \approx 10^{-4}$ Pa, as shown in \cref{G_complex}b. Scaling arguments show that the bending modulus and plateau modulus are related as $\kappa \sim G_0 R^3$, where $R$ is the equivalent radius of vesicle. For a typical vesicle size $R =$ 10 $\mu$m, the bending modulus can be estimated as $\kappa \approx 10^{-19} J$, or $\kappa \approx 24$ $k_BT$ which is close to the experimentally measured value for DOPC lipid vesicles in this work \cite{kumar2020conformational,dahl2016experimental}. 

In this letter, we directly observe the relaxation of highly deformed vesicles in quiescent solution. Our results show that vesicles dissipate stress via two distinct and well-separated time scales, with a fast and a slow time scale attributed to the relaxation of bending fluctuation modes and surface tension-dominated modes, respectively. Broadly speaking, these results show how the interplay between vesicle excess area, bending forces, and surface tension yields a complex relaxation behavior that has not been previously observed in tethers extruded from quasi-spherical vesicles \cite{brochard2006hydrodynamic,rossier2003giant}. These results highlight the use of the Stokes trap in observing the dynamic behavior of vesicle shape under precisely defined flows. This methodology of combining gentle flow-based trapping with fluorescence microscopy to induce high-levels of membrane deformation will open new avenues in understanding the dynamics of other membrane-bound particles such as polymersomes, capsules, and living cells without the need for micropipettes or external manipulation of membranes.

We thank Noah Hopkins for help in analyzing experimental data. This work was supported by National Science Foundation by Award $\#$ NSF CBET 1704668.

%\bibliography{refs}


\end{document}